\documentclass[reqno,centertags,12pt]{amsart}

\usepackage[utf8]{inputenc}
\usepackage[T1]{fontenc}
\usepackage[american]{babel}
\usepackage{iftex}
\ifXeTeX
\usepackage[babel=true,expansion=false,protrusion=alltext-nott,final]{microtype}
\else
\usepackage[babel=true,expansion=alltext,protrusion=alltext-nott,final]{microtype}
\fi
\usepackage[sb,type1]{libertine}
\usepackage[varqu,varl]{inconsolata}
\usepackage[libertine]{newtxmath}
\useosf
\usepackage{textcomp}
\usepackage{dsfont}

\usepackage{amsthm,mathtools}
\usepackage{graphicx}
\usepackage{float}
\usepackage{enumitem,multirow}
\usepackage{needspace}
\usepackage[margin={3cm},marginparwidth={2.5cm}]{geometry}
\usepackage[svgnames]{xcolor}
\usepackage[breaklinks=true,pagebackref]{hyperref}

\hypersetup{
  pdftitle={The sharp fermionic Hardy constant in dimension two},
  pdfauthor={Marvin R. Schulz},
  pdfsubject={Mathematical physics and spectral inequalities},
  pdfkeywords={Hardy inequalities, fermions, two dimensions, semiclassical asymptotics}
}

\newcommand{\N}{{\mathbb{N}}}
\newcommand{\R}{{\mathbb{R}}}

\newcommand{\C}{{\mathbb{C}}}

\newcommand{\re}{\mathrm{Re}}

\newcommand{\abs}[1]{\left|#1\right|}

\newcommand{\1}{\mathbf 1}
\newcommand{\dd}{\,d}

\allowdisplaybreaks

\newtheorem{theorem}{Theorem}[section]
\newtheorem{proposition}[theorem]{Proposition}
\newtheorem{lemma}[theorem]{Lemma}

\theoremstyle{definition}

\theoremstyle{remark}
\newtheorem{remark}[theorem]{Remark}

\numberwithin{equation}{section}

\newcounter{smalllist}

\title[Sharp 2D fermionic Hardy Inequality]{Sharp constant in 2D critical fermionic Hardy inequality}

\author{Marvin R. Schulz}
\address{Department of Mathematical Sciences, University of Copenhagen, Universitetsparken 5, 2100 Copenhagen, Denmark}
\email{masc@math.ku.dk}

\date{\today, version \jobname}


\begin{document}

\begin{abstract}
We study the large-$N$ behavior of the optimal constant $\kappa_N$ in the two-dimensional fermionic many-particle Hardy inequality
\[
\sum_{i=1}^N -\Delta_i
\geq
\kappa_N^{(2)}\sum_{1\leq i<j\leq N}\frac{1}{|x_i-x_j|^2},
\]
understood as a quadratic-form inequality on antisymmetric wave functions. We prove that
\[
\lim_{N\to\infty}(\log N)\kappa_N^{(2)}=4,
\]
thereby proving a conjecture of Frank, Hoffmann-Ostenhof, Laptev, and Solovej~\cite{FHOLS}. The proof combines a dyadic Fefferman--de la Llave representation of the Coulomb kernel with separate estimates for squares of low and high particle occupation. A trace inequality controls the low-occupation regime, while sums of Neumann eigenvalues yield the sharp bound in the high-occupation regime.
\end{abstract}

\maketitle

{\hypersetup{linkcolor=black}
\tableofcontents}
\section{Fermionic Hardy Inequality}
Consider $N \ge 2$ fermionic particles in two-dimensional space with positions $X=(x_1,\ldots,x_N)$ such that $X\in\R^{2N}$. We denote by $\Psi:\R^{2N}\to \C$ the corresponding antisymmetric state and define 
\[
T[\Psi]=\sum_{i=1}^N\int_{\R^{2N}}|\nabla_i\Psi(X)|^2\dd X,
 \qquad V(X)=\sum_{1\le i<j\le N}|x_i-x_j|^{-2}.
\]
We aim to determine the following Hardy constant
\begin{equation}\label{eq:kappa}
 \kappa_N^{(2)}
 =\inf_{\substack{0\ne\Psi\in\dot H^1(\R^{2N})\\
                         \Psi\ \mathrm{antisymmetric}}}
                    \frac{T[\Psi]}{\int V|\Psi|^2\dd X}.
\end{equation}
Here the Sobolev space $\dot H^1(\R^{2N})$ is the completion of $C_c^\infty(\R^{2N})$ in the gradient norm. Following
\cite[Theorem~2.8]{HOHLT} and \cite[Appendix B, Proposition 15]{FHOLS} we have
\begin{equation}\label{eq:known}
 \kappa_N^{(2)}\ge\frac4N,
 \qquad
 \limsup_{N\to\infty}(\log N)\kappa_N^{(2)}\le4.
\end{equation}
Frank, Hoffmann-Ostenhof, Laptev, and Solovej
\cite[Theorem~1]{FHOLS} established sharp large-$N$ asymptotics for the corresponding constants with Riesz kernels $|x_i-x_j|^{-2s}$ in the subcritical range $0<s<d/2$, $s\leq 1$. Their proof adapts the strategy of Lieb and Yau in their analysis of Chandrasekhar's theory of stellar collapse~\cite{LiebYau}.

In \cite{FHOLS} the authors pose the question wether or not $4$ is the sharp constant. We prove that this is indeed the case by proving the following theorem
\begin{theorem}\label{thm:sharp}
There is a constant $C>0$ such that for every $N\ge2$ and every
antisymmetric $\Psi\in\dot H^1(\R^{2N})$,
\begin{equation}\label{eq:main-lower}
 \int V|\Psi|^2\dd X
 \le\left(\frac{\log N}{4}+C\right)T[\Psi],
\end{equation}
and consequently
\begin{equation}\label{eq:sharp-limit}
 \lim_{N\to\infty}(\log N)\kappa_N^{(2)}=4.
\end{equation}
\end{theorem}

\paragraph{\textbf{Structure of the paper:} }
In Section~\ref{sec:sharp}, we give a Fefferman--de la Llave-type representation of the inverse-square potential based on a dyadic decomposition of $\mathbb{R}^2$ into squares.

In Section~\ref{sec:proof-of-proposition}, we divide the argument into \textit{low- and high-occupation regimes}, according to the number of particles in each square. In the low-occupation regime, we prove a Hardy inequality for nearest-neighbour sets using a trace inequality. In the high-occupation regime, we derive a sharp bound in terms of sums of Neumann eigenvalues on the relevant squares. These two estimates are given in Lemmas~\ref{lem:grid-low} and~\ref{lem:grid-high}, respectively. Together, they imply Proposition~\ref{prop:dyadic-sum-bound}, which, combined with the Fefferman--de la Llave representation, yields the main result.

\section{Main Theorem}\label{sec:problem}
We consider $N\ge 2$ fermionic particles at positions $X=(x_1,\ldots,x_N)\in(\R^2)^N$. We denote by $\Psi:\R^{2N} \to \C$ the corresponding state which is antisymmetric under exchange of particles. We define
\[
 T[\Psi]=\sum_{i=1}^N\int_{\R^{2N}}|\nabla_i\Psi(X)|^2\dd X,
 \qquad V(X)=\sum_{1\le i<j\le N}|x_i-x_j|^{-2}.
\]
We define the following Hardy constant 
\begin{equation}\label{eq:kappa}
 \kappa_N^{(2)}
 =\inf_{\substack{0\ne\Psi\in\dot H^1(\R^{2N})\\
                         \Psi\ \mathrm{antisymmetric}}}
                    \frac{T[\Psi]}{\int V|\Psi|^2\dd X}.
\end{equation}

Here the Sobolev space $\dot H^1(\R^{2N})$ is the completion of
$C_c^\infty(\R^{2N})$ in the gradient norm. 

Following \cite[Appendix B, Proposition 15]{FHOLS} and we have
\begin{equation}\label{eq:known}
 \kappa_N^{(2)}\ge\frac4N,
 \qquad
 \limsup_{N\to\infty}(\log N)\kappa_N^{(2)}\le4.
\end{equation}
The question whether or not the constant is sharp is asked by the authors and left open. The aim of the note is to prove 
\begin{equation}\label{eq:question}
 \lim_{N\to\infty}(\log N)\kappa_N^{(2)}=4.
\end{equation}

The strategy of the proof is to use a dyading tiling of $\R^2$ with a squares. Using these squares we apply a  Fefferman Fefferman de la Llave representation of the Hardy potential and estimate contributions for cases in which few particles are in the participating squares (low occupation) and cases in which many particles are in the participating squares (high occupation) independently. 

For the low occupation squares a nearest-neigbour Hardy inequality obtained from the underlying antisymmetrie and a trace estimate controls this regime and yields a contribution which is negligible in the large $N$ limit. 
For the high occupied squares we obtain sharp leading constant from Neumann eigenvalues of the participating squares.

\section{Proof of Main Theorem}\label{sec:sharp}
Fefferman and de la Llave \cite{FdL} represented the three-dimensional
Coulomb kernel in terms of overlaps of balls. Hainzl and Seiringer
\cite[Theorem~1]{HS} extended this representation to more general radial
kernels. We use a square-grid analogue for $|x|^{-2}$ in two dimensions.

For each side length $\ell>0$, angle $\theta\in[0,2\pi)$, and shift
$z\in[0,\ell)^2$, define the set of squares
\[
 \Gamma(\ell,\theta,z)=\bigl\{R_\theta\bigl(z+\ell(j+[0,1)^2)\bigr):
                         j\in\mathbb Z^2\bigr\},
\]
where $R_\theta$ denotes rotation through $\theta$. For a square $Q$
and a configuration $X$, let
\begin{equation}
 n_Q(X)=\#\{i:x_i\in Q\}
\end{equation}
denote the number of particles in $Q$. We integrate over the side length
and average over rotations and shifts with respect to the measure
\begin{equation}
 d\mu(\ell,\theta,z)=\frac{d\ell}{\ell^3}\,
                     \frac{d\theta}{2\pi}\,\frac{dz}{\ell^2}.
\end{equation}
In the following integrals, $\Gamma$ denotes $\Gamma(\ell,\theta,z)$.
Appendix
\ref{app:square-representation} shows that for every configuration $X$
of distinct points,
\begin{equation}\label{eq:square-rep}
 V(X)=\pi\int \sum_{Q\in \Gamma} n_Q(X)(n_Q(X)-1)\dd\mu.
\end{equation}

\begin{figure}[!htbp]
 \centering
 \includegraphics[width=\linewidth]{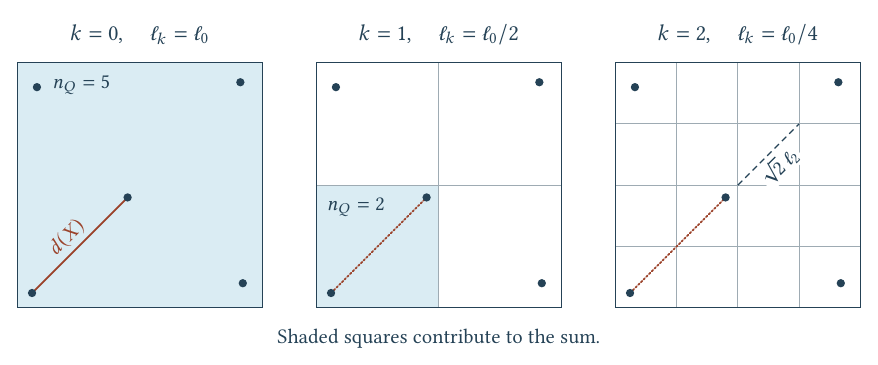}
 \caption{Example of $5$ particles and the first three subdivision. At level $k=2$, the square diameter $\sqrt2\,\ell_2$ is smaller than the minimum pair distance $d(X)$, so all summands at this level in \eqref{eq:dyadic-sum} vanish.}
 \label{fig:dyadic-grid}
\end{figure}

Fix a coarsest side length $\ell_0>0$, an angle $\theta\in[0,2\pi)$,
and a shift $z\in[0,\ell_0)^2$. Dyadic subdivision of the squares in
$\Gamma(\ell_0,\theta,z)$ gives the grids
\[
 \ell_k=2^{-k}\ell_0,\qquad
 \Gamma_k=\bigl\{R_\theta\bigl(z+\ell_k(j+[0,1)^2)\bigr)
                       :j\in\mathbb Z^2\bigr\},
 \qquad k=0,1,2,\ldots.
\]
With $z$ and $\theta$ fixed, each square of $\Gamma_k$ is the disjoint
union of four squares of $\Gamma_{k+1}$. For distinct points $X$, set
\begin{equation}\label{eq:dyadic-sum}
 S_{\ell_0}(X)=\sum_{k\ge0}\sum_{Q\in\Gamma_k}
                  \frac{n_Q(X)(n_Q(X)-1)}{|Q|}.
\end{equation}
The notation suppresses the dependence on $\theta$ and $z$. For all
sufficiently large $k$,
\begin{equation}
 \sqrt2\,\ell_k < d(X),\qquad d(X)=\min_{i<j}|x_i-x_j|.
\end{equation}
At these levels, each $Q\in\Gamma_k$ has $n_Q(X)\in\{0,1\}$, so its
summand vanishes. At each of the finitely many remaining levels, at
most $\lfloor N/2\rfloor$ squares
can contain two or more particles, since the squares at that level are
disjoint. Thus the sum has finitely many nonzero terms, as illustrated
in Figure~\ref{fig:dyadic-grid}.

For $L>0$, define the truncated potential by restricting the
side-length integral in \eqref{eq:square-rep} to $0<\ell\le L$,
\begin{equation}\label{eq:truncated-potential}
 V_L(X)=\pi\int_{0<\ell\le L}
   \sum_{Q\in\Gamma(\ell,\theta,z)}n_Q(X)(n_Q(X)-1)\,
   d\mu(\ell,\theta,z).
\end{equation}
\begin{lemma}[Dyadic representation]\label{lem:truncated-representation}
For every $L>0$ and every configuration $X$ of distinct points,
\begin{equation}\label{eq:grid-average}
 V_L(X)=\pi\int_{L/2<\ell_0\le L} S_{\ell_0}(X)\,
                     \ell_0^2\dd\mu(\ell_0,\theta,z).
\end{equation}
\end{lemma}
\begin{remark}
The proof is given in Appendix~\ref{app:truncated-representation}.
\end{remark}
For distinct points $X$, the nonnegativity of the integrand and
\eqref{eq:square-rep} give $V_L(X)\uparrow V(X)$ as $L\to\infty$ by
monotone convergence. Lemma~\ref{lem:truncated-representation} reduces
Theorem~\ref{thm:sharp} to an estimate uniform in the grid parameters.

\begin{proposition}\label{prop:dyadic-sum-bound}
There is a constant $C_0>0$ independent of $\ell_0,\theta$ and $z$ such that, for every $N\ge2$ and every smooth compactly supported antisymmetric $\Psi$,
\begin{equation}\label{eq:dyadic-sum-bound}
 \int S_{\ell_0}|\Psi|^2\dd X
 \le \left(\frac{\log N}{2\pi\log4}+C_0\right)T[\Psi].
\end{equation}
\end{proposition}
\begin{remark}
The proof of Proposition~\ref{prop:dyadic-sum-bound} is the content of Section~\ref{sec:proof-of-proposition}.
\end{remark}
Assuming Proposition~\ref{prop:dyadic-sum-bound} for now we can proof the main theorem.
\begin{proof}[Proof of Theorem \ref{thm:sharp}]
First suppose that $\Psi$ is smooth, compactly supported, and
antisymmetric. Using \eqref{eq:grid-average} and Tonelli's theorem, we
integrate \eqref{eq:dyadic-sum-bound} over the grid parameters to obtain
\[
 \begin{aligned}
 \int V_L|\Psi|^2\dd X
 &=\pi\int_{L/2<\ell_0\le L}
    \left(\int S_{\ell_0}|\Psi|^2\dd X\right)
    \ell_0^2\dd\mu(\ell_0,\theta,z)\\
 &\le\pi\left(\frac{\log N}{2\pi\log4}+C_0\right)T[\Psi]
    \int_{L/2<\ell_0\le L}\ell_0^2\dd\mu(\ell_0,\theta,z).
 \end{aligned}
\]
The integral over the grid parameters is
\[
 \int_{L/2<\ell_0\le L}\ell_0^2\dd\mu(\ell_0,\theta,z)
 =\int_{L/2}^{L}\frac{\dd\ell_0}{\ell_0}
   \left(\int_0^{2\pi}\frac{\dd\theta}{2\pi}\right)
   \left(\int_{[0,\ell_0)^2}\frac{\dd z}{\ell_0^2}\right)
 =\int_{L/2}^{L}\frac{\dd\ell_0}{\ell_0}=\log 2.
\]
Letting $L\to\infty$ and applying monotone convergence to the integral
in $X$ yields \eqref{eq:main-lower} with $C=\pi(\log2)C_0$.

For general antisymmetric $\Psi\in\dot H^1(\R^{2N})$, approximate in the gradient norm by smooth compactly supported functions and project it onto antisymmetic functions. The Sobolev inequality gives convergence in $L^{2N/(N-1)}(\R^{2N})$, so a subsequence converges almost everywhere. Fatou's lemma then extends \eqref{eq:main-lower} to $\Psi$. It follows that $\kappa_N^{(2)}\ge 4/(\log N+4C)$. Together with the upper bound in \eqref{eq:known}, this proves \eqref{eq:sharp-limit}.
\end{proof}

\Needspace{11\baselineskip}
\section{High and Low Occupation Regimes}\label{sec:proof-of-proposition}

Throughout this section, fix the grid parameters $\ell_0>0$,
$\theta\in[0,2\pi)$, and $z\in[0,\ell_0)^2$, and let $\Psi$ be as in
Proposition~\ref{prop:dyadic-sum-bound}. For an integer $M\ge1$, define
the low- and high-occupancy parts of \eqref{eq:dyadic-sum} by
\begin{align}
 S_{\ell_0,\mathrm{low}}^{(M)}(X)
 &=\sum_{k\ge0}\sum_{\substack{Q\in\Gamma_k\\n_Q(X)\le M}}
        \frac{n_Q(X)(n_Q(X)-1)}{|Q|},\notag\\
 S_{\ell_0,\mathrm{high}}^{(M)}(X)
 &=\sum_{k\ge0}\sum_{\substack{Q\in\Gamma_k\\n_Q(X)>M}}
        \frac{n_Q(X)(n_Q(X)-1)}{|Q|}.
 \label{eq:high-occupancy-sum}
\end{align}
Thus
\[
 S_{\ell_0}(X)=S_{\ell_0,\mathrm{low}}^{(M)}(X)
                    +S_{\ell_0,\mathrm{high}}^{(M)}(X).
\]
We first state estimates of the low- and high-occupancy parts independently as Lemma~\ref{lem:grid-low} and Lemma~\ref{lem:grid-high} and show how they conclude Proposition~\ref{prop:dyadic-sum-bound}
\begin{lemma}[Low occupancies]\label{lem:grid-low}
For every integer $M\ge1$,
\begin{equation}\label{eq:grid-low}
 \int S_{\ell_0,\mathrm{low}}^{(M)}(X)|\Psi(X)|^2\dd X
 \le \frac{64}{3}(M-1)T[\Psi].
\end{equation}
\end{lemma}

\begin{lemma}[High occupancies]\label{lem:grid-high}
Let $m>32\pi$ be an integer and set $M=16m$. Then
\begin{equation}\label{eq:grid-high}
 \int S_{\ell_0,\mathrm{high}}^{(M)}(X)|\Psi(X)|^2\dd X
 \le \frac{\log N+11}{2\pi\log4}\,T[\Psi].
\end{equation}
\end{lemma}

\begin{proof}[Proof of Proposition \ref{prop:dyadic-sum-bound}]
Fix an integer $m>32\pi$ independently of $N$ and take $M=16m$ in
both lemmas. Adding \eqref{eq:grid-low} and \eqref{eq:grid-high} and
using the decomposition of $S_{\ell_0}$ proves
\eqref{eq:dyadic-sum-bound} with
$C_0=\frac{64}{3}(M-1)+\frac{11}{2\pi\log4}$. Both bounds are uniform in
$\ell_0$, $\theta$, and $z$.
\end{proof}

\subsection{Low occupancies}
In this subsection we prove Lemma~\ref{lem:grid-low} which corresponds to the part of the potential generated by squares of low occupation. We first state a Hardy inequality for nearest neighbors that is uniform in the number of particles as Proposition~\ref{prop:nearest-neighbour} and show that this inequality concludes Lemma~\ref{lem:grid-low}.

For a configuration $X$ of distinct particles, we define the distance to the nearest neighbor as
\[
 d_i(X)=\min_{j\ne i}|x_i-x_j|>0.
\]
and we claim for every $N\ge2$ and every scalar antisymmetric
$\Psi\in\dot H^1(\R^{2N})$,
\begin{equation}\label{eq:claim:nearest-neighbour}
 \sum_{i=1}^N\int\frac{|\Psi(X)|^2}{d_i(X)^2}\dd X
 \le8T[\Psi].
\end{equation}
We prove Lemma~\ref{lem:grid-low} using \eqref{eq:claim:nearest-neighbour}
and then establish the claim in Proposition~\ref{prop:nearest-neighbour}.

\begin{proof}[Proof of Lemma \ref{lem:grid-low}]
Fix a configuration $X$ of distinct particles. Let $\widetilde{Q}_{k,i}$
be the unique square in $\Gamma_k$ containing $x_i$. For $M\ge1$ and
$k\ge0$, we have
\[
 \begin{aligned}
 \sum_{\substack{Q\in\Gamma_k\\n_Q(X)\le M}}
       n_Q(X)\bigl(n_Q(X)-1\bigr)
 &=\sum_{\substack{Q\in\Gamma_k\\2\le n_Q(X)\le M}}
       \sum_{i:\,x_i\in Q}\bigl(n_Q(X)-1\bigr)\\
 &=\sum_{i=1}^N \1_{\{2\le n_{\widetilde{Q}_{k,i}}(X)\le M\}}
       \bigl(n_{\widetilde{Q}_{k,i}}(X)-1\bigr)\\
 &\le (M-1)\sum_{i=1}^N\1_{\{2\le n_{\widetilde{Q}_{k,i}}(X)\le M\}}.
 \end{aligned}
\]
If $n_{\widetilde{Q}_{k,i}}(X)\ge2$, there is another particle $x_j$ in
the same square, so $d_i(X)\le |x_i-x_j|\le\sqrt2\,\ell_k$. Hence
\[
 \sum_{\substack{Q\in\Gamma_k\\n_Q(X)\le M}}
       n_Q(X)\bigl(n_Q(X)-1\bigr)
 \le (M-1)\sum_{i=1}^N\1_{\{d_i(X)\le\sqrt2\,\ell_k\}}.
\]
Since $|Q|=\ell_k^2$ for $Q\in\Gamma_k$, multiplying by $\ell_k^{-2}$
and summing over $k$ gives
\begin{equation}\label{eq:up_to_dyadic_sum}
 S_{\ell_0,\mathrm{low}}^{(M)}(X)
 \le (M-1)\sum_{i=1}^N
       \sum_{\substack{k\ge0\\\ell_k\ge d_i(X)/\sqrt2}}\ell_k^{-2}.
\end{equation}
For each $i$ for which the inner sum is nonempty, define
\[
 K_i=\max\{k\in\N_0:\sqrt2\,\ell_k\ge d_i(X)\}.
\]
This maximum is finite because $d_i(X)>0$ and $\ell_k\to0$.
Since $\ell_k=2^{K_i-k}\ell_{K_i}$ for $0\le k\le K_i$, we have
\begin{equation}\label{eq:geometric_series}
 \begin{aligned}
 \sum_{\substack{k\ge0\\\ell_k\ge d_i(X)/\sqrt2}}\ell_k^{-2}
 &=\ell_{K_i}^{-2}\sum_{k=0}^{K_i}\left(\frac14\right)^{K_i-k}\\
 &\le \ell_{K_i}^{-2}\sum_{n=0}^{\infty}\left(\frac14\right)^n
 =\frac43\ell_{K_i}^{-2}
 \le \frac{8}{3d_i(X)^2}.
 \end{aligned}
\end{equation}
The last inequality uses $\ell_{K_i}\ge d_i(X)/\sqrt2$. If the inner
sum is empty, it is zero. Combining \eqref{eq:up_to_dyadic_sum} and
\eqref{eq:geometric_series} gives
\begin{equation}\label{eq:low_oc_nn}
 S_{\ell_0,\mathrm{low}}^{(M)}(X)
 \le \frac83(M-1)\sum_{i=1}^N d_i(X)^{-2}.
\end{equation}
Integrating \eqref{eq:low_oc_nn} against $|\Psi|^2$ and applying
\eqref{eq:claim:nearest-neighbour} gives
\begin{equation}
 \int S_{\ell_0,\mathrm{low}}^{(M)}(X)|\Psi(X)|^2\dd X
 \le \frac83(M-1)\sum_{i=1}^N\int\frac{|\Psi(X)|^2}{d_i(X)^2}\dd X
 \le \frac{64}{3}(M-1)T[\Psi].
\end{equation}
\end{proof}
\begin{proposition}[Nearest-neighbor estimate]\label{prop:nearest-neighbour}
For every $N\ge2$ and every scalar antisymmetric
$\Psi\in\dot H^1(\R^{2N})$,
\begin{equation}\label{eq:nearest-neighbour}
 \sum_{i=1}^N\int\frac{|\Psi(X)|^2}{d_i(X)^2}\dd X
 \le8T[\Psi].
\end{equation}
\end{proposition}

\begin{proof}[Proof of Proposition~\ref{prop:nearest-neighbour}] The idea is similar to \cite[Section~5 and Appendix~A]{HSVlines}, where a trace estimates on conical boundaries are combined with Hardy inequalities to control localization terms.

First let $\Psi$ be smooth, compactly supported, and antisymmetric. We decompose configuration space according to the nearest neighbour of each particle. For distinct indices $i,j\in\{1,\ldots,N\}$, let
\[
 D_{ij}=\{X\in\R^{2N}:\abs{x_i-x_j}<\abs{x_i-x_k}
                   \text{ for all }k\ne i,j\}.
\]
Thus $D_{ij}$ consists of the configurations in which $x_j$ is the unique nearest neighbour of $x_i$. We use two properties of these sets.
\begin{enumerate}
 \item[i)] For each fixed $i$, the sets $D_{ij}$, $j\ne i$, are
 pairwise disjoint and cover $\R^{2N}$ up to a null set.
 \item[ii)] For every configuration $X$ of distinct points and every
 fixed $i$, at most five particles have $x_i$ as their unique nearest
 neighbour. In other words,
 \begin{equation}\label{eq:nn-multiplicity}
  \#\{j\ne i:X\in D_{ji}\}\le5.
 \end{equation}
\end{enumerate}

\textit{Claim (i).}
Disjointness follows from the strict inequalities in the definition
of $D_{ij}$. If $X\notin\bigcup_{j\ne i}D_{ij}$, then the minimum
distance from $x_i$ to the other particles is attained at two
distinct indices $\ell,m\ne i$, so
\[
 \abs{x_i-x_\ell}=\abs{x_i-x_m}.
\]
For a fixed pair $\ell,m$, fixing all coordinates except $x_i$ with
$x_\ell\ne x_m$ confines $x_i$ to an affine line. By Fubini's theorem, these configurations form a null set. The configurations with $x_\ell=x_m$ also form a null set, since fixing all coordinates except
$x_\ell$ confines $x_\ell$ to the single point $x_m$. Taking the finite union over pairs $\ell,m$ proves (i).

\textit{Claim (ii).}
Fix a configuration $X$ of distinct points and an index $i$. Let
$\ell,k\ne i$ be distinct indices with $X\in D_{\ell i}\cap D_{ki}$.
Interchanging $\ell$ and $k$ if necessary, we may assume that
\[
 0<t:=\abs{x_\ell-x_i}\le\abs{x_k-x_i}=:s.
\]
Since $X\in D_{ki}$, we have $\abs{x_k-x_\ell}>s$. For the angle
$\alpha$ between $x_\ell-x_i$ and $x_k-x_i$, the cosine rule gives
\[
 t^2+s^2-2ts\cos\alpha=\abs{x_k-x_\ell}^2>s^2.
\]
Hence $\cos\alpha<t/(2s)\le1/2$ and $\alpha>\pi/3$. Assume that there are more than five particles that have $x_i$ as their unique nearest neighbour then the angular distance between to of them must be at most $\pi/3$ which is contradiction to $\alpha>\pi/3$. This proves (ii).\\

For each $i$, we have $d_i(X)=|x_i-x_j|$ on $D_{ij}$, so (i) gives
\begin{equation}\label{eq:nn-decomposition}
 \sum_{i=1}^N\int\frac{|\Psi(X)|^2}{d_i(X)^2}\dd X
 =\sum_{i\ne j}\int_{D_{ij}}
   \frac{|\Psi(X)|^2}{|x_i-x_j|^2}\dd X.
\end{equation}

To estimate a term on the right we use relative coordinates for each fixed pair $(i,j)$. Denote by $R=(x_i+x_j)/2$ the midpoint and by $r=x_i-x_j$ the relative position between $x_i$ and $x_j$.
We then can write $X(r) \in \R^{2N}$ for any configuration with $i$th coordinate $R+r/2$, $j$th coordinate $R-r/2$ and define $u\colon\R^2\to\mathbb C$ by
\[
 u(r)=\Psi(X(r)).
\]
Replacing $r$ by $-r$ exchanges the $i$th and $j$th coordinates, so
$u(-r)=-u(r)$ by antisymmetry. The chain rule gives
\[
 \nabla_r u(r)=\tfrac12(\nabla_i\Psi-\nabla_j\Psi)(X(r)).
\]
\paragraph{\textbf{Poincar\'e inequality on the circle}}
For fixed $\rho>0$, the function
\[
 f_\rho(\theta)=u\bigl(\rho(\cos\theta,\sin\theta)\bigr)
\]
is smooth and $2\pi$-periodic. Since $f_\rho(\theta+\pi)=-f_\rho(\theta)$,
it has zero mean. For the Fourier coefficients
$c_n=(2\pi)^{-1}\int_0^{2\pi} f_\rho(\theta)e^{-in\theta}\dd\theta$,
this gives $c_0=0$. The
$n$th Fourier coefficient of $\partial_\theta f_\rho$ is $i n c_n$.
Following Parseval's identity we have $f_\rho,\partial_\theta f_\rho\in L^2(0,2\pi)$
and find
\begin{equation}\label{eq:angular_mean_van}
 \int_0^{2\pi}|f_\rho|^2\dd\theta
 =2\pi\sum_{n\ne0}|c_n|^2
 \le 2\pi\sum_{n\ne0}n^2|c_n|^2
 =\int_0^{2\pi}|\partial_\theta f_\rho|^2\dd\theta.
\end{equation}
Integrating against $\dd\rho/\rho$ over $0<\rho<s$ and using
$\dd r=\rho\,\dd\rho\dd\theta$ and
$|\nabla_r u|^2=|\partial_\rho u|^2+\rho^{-2}|\partial_\theta u|^2$,
we obtain, for every $s>0$,
\begin{equation}\label{eq:nn-disk-hardy}
 \int_{|r|<s}\frac{|u(r)|^2}{|r|^2}\dd r
 \le \int_{|r|<s}\left|\frac{1}{\rho}\partial_\theta u\right|^2\dd r
 \le \int_{|r|<s}|\nabla_r u|^2\dd r.
\end{equation}

Set $y_k=x_k-R$ for $k\ne i,j$. The condition $X(r)\in D_{ij}$
restricts $r$ to the domain
\[
 \Omega=\{r\in\R^2: |r|<|r/2-y_k|\text{ for all }k\ne i,j\}.
\]
For each $k\ne i,j$,
\[
 |r|^2<|r/2-y_k|^2
 \quad\Longleftrightarrow\quad
 \tfrac34|r|^2+r\cdot y_k<|y_k|^2
 \quad\Longleftrightarrow\quad
 \left|r+\tfrac23y_k\right|^2<\tfrac{16}{9}|y_k|^2.
\]
Thus
\[
 \Omega=\bigcap_{k\ne i,j}B_{\frac43|y_k|}\!\left(-\tfrac23y_k\right),
\]
where $B_s(c)=\{r\in\R^2:|r-c|<s\}$ for $s\ge0$. For $N\ge3$, set
$a=\frac23\min_{k\ne i,j}|y_k|$. We may assume $a>0$, since
$\Omega=\varnothing$ when $a=0$. Each disc has center
$c_k=-\tfrac23y_k$ and radius $2|c_k|$, so it contains $B_a(0)$ by the
triangle inequality. For $r\in\Omega$, choose $k$ with $|c_k|=a$. Then
\begin{equation}
    \abs{r}\leq \abs{r-c_k}+\abs{c_k} < 2a +a=3a
\end{equation}
and consequently
\begin{equation}\label{eq:nn-domain-bounds}
 B_a(0)\subset\Omega\subset B_{3a}(0).
\end{equation}
In Figure~\ref{fig:nearest-neighbour-domain} we illustrate the set $\Omega$.
\begin{figure}[H]
 \centering
 \includegraphics[width=\linewidth]{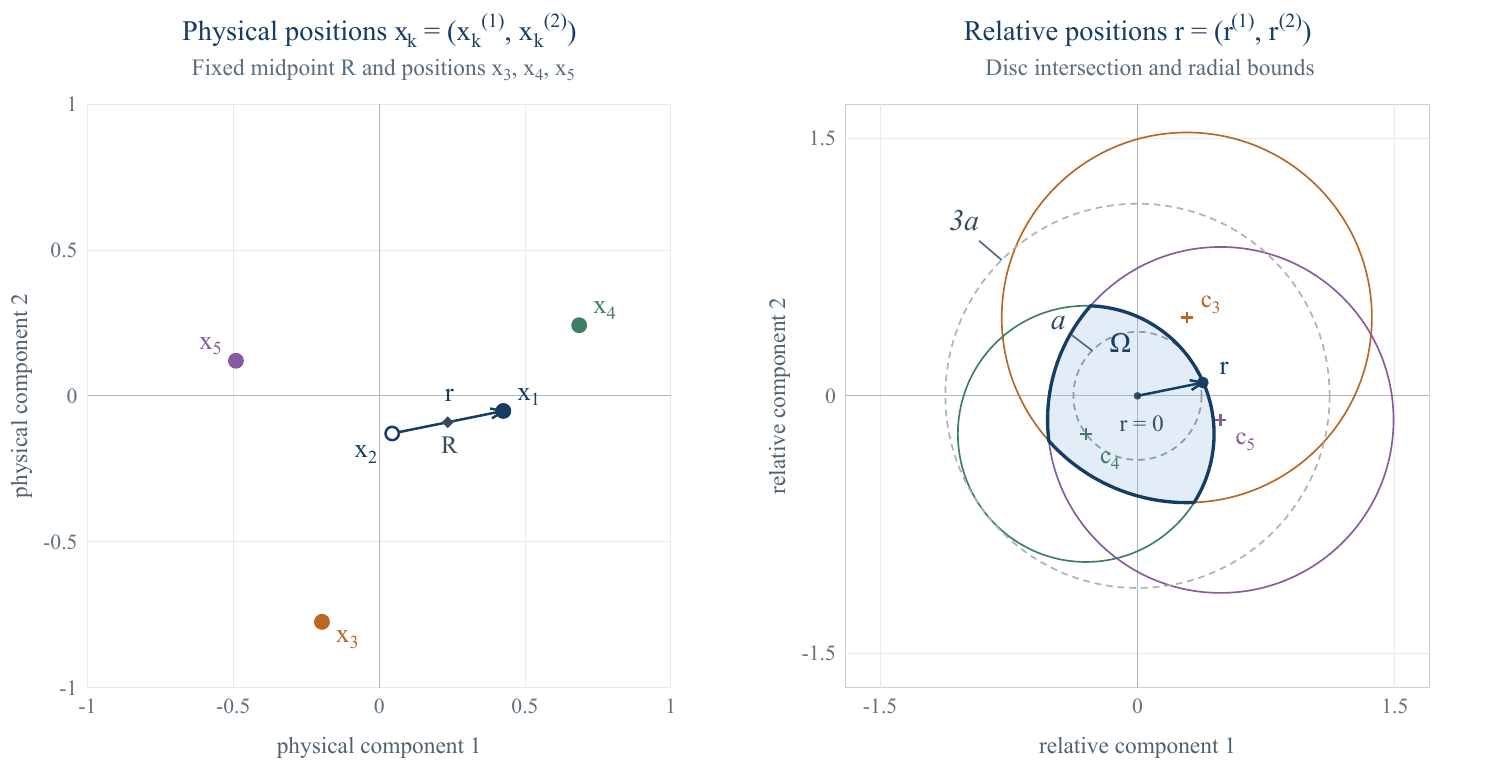}
 \caption{The set $\Omega$ for $N=5$ and $(i,j)=(1,2)$. On the left, the midpoint $R$ and the positions $x_3,x_4,x_5$ are fixed the relative coordinate $r$ and center of mass $R$ are indicated.
 On the right, the shaded region $\Omega$ is the intersection of the three discs with centres $c_k=-2y_k/3$ and radii $2|c_k|$. The dashed circles bound $B_a(0)$ and $B_{3a}(0)$.}
 \label{fig:nearest-neighbour-domain}
\end{figure}

Write $B_a=B_a(0)$. We split the Hardy integral as
\begin{equation}
 \int_\Omega\frac{|u|^2}{|r|^2}\dd r
 =\int_{B_a}\frac{|u|^2}{|r|^2}\dd r
  +\int_{\Omega\setminus B_a}\frac{|u|^2}{|r|^2}\dd r.
\end{equation}
Inequality \eqref{eq:nn-disk-hardy} with $s=a$ controls the first term
on the right-hand side.

Note that $\Omega$ is convex by construction as overlap of disks and contains the origin. We conclude
\begin{equation}
 \Omega\setminus B_a
 =\bigl\{\rho(\cos\theta,\sin\theta)
        \,\bigm|\, 0\le\theta<2\pi,\quad a\le\rho<b(\theta)\bigr\},
\end{equation}
where $b(\theta)$ is a parametrization of $\partial\Omega$. We have $a\le b(\theta)\le3a$ by
\eqref{eq:nn-domain-bounds}.

For $a\le t<b(\theta)$, radial integration from $\partial B_a$ gives
\[
 u(t,\theta)=u(a,\theta)+\int_a^t\partial_\rho u(\rho,\theta)\dd\rho.
\]
Using $2|z||w|\le\tfrac13|z|^2+3|w|^2$, we have
$|z+w|^2\le\tfrac43|z|^2+4|w|^2$. Applying this to the radial identity
yields
\[
 |u(t,\theta)|^2
 \le \frac43|u(a,\theta)|^2
   +4\left|\int_a^t\partial_\rho u(\rho,\theta)\dd\rho\right|^2.
\]
For the integral term, write $\partial_\rho u$
as $(\sqrt{\rho}\,\partial_\rho u)\rho^{-1/2}$. By Cauchy--Schwarz,
\[
 \begin{aligned}
 \left|\int_a^t\partial_\rho u(\rho,\theta)\dd\rho\right|^2
 &\le \left(\int_a^t|\partial_\rho u(\rho,\theta)|^2\rho\,\dd\rho\right)
      \left(\int_a^t\frac{\dd\rho}{\rho}\right)\\
 &=\log(t/a)\int_a^t|\partial_\rho u(\rho,\theta)|^2\rho\,\dd\rho.
 \end{aligned}
\]
Extending the energy integral to $b(\theta)$ and substituting the bound
above, we obtain
\[
 |u(t,\theta)|^2
 \le \frac43|u(a,\theta)|^2
   +4\log(t/a)\int_a^{b(\theta)}
                |\partial_\rho u(\rho,\theta)|^2\rho\,\dd\rho.
\]
Integrating with $\dd t/t$ from $a$ to $b(\theta)$ gives the
coefficients $\tfrac43\log(b(\theta)/a)$ and
$2\log^2(b(\theta)/a)$, respectively.
Since $b(\theta)/a\le3$, integration in $\theta$ gives
\begin{equation}\label{eq:nn-outer-hardy}
 \int_{\Omega\setminus B_a}\frac{|u|^2}{|r|^2}\dd r
 \le \frac43\log3\int_0^{2\pi}|u(a,\theta)|^2\dd\theta
       +2(\log3)^2\int_{\Omega\setminus B_a}|\partial_\rho u|^2\dd r.
\end{equation}

To bound the trace term in \eqref{eq:nn-outer-hardy}, use $u(0)=0$ by
oddness. For fixed $\theta$, the fundamental theorem of calculus gives
\[
 |u(a,\theta)|^2
 =\int_0^a\partial_\rho |u(\rho,\theta)|^2\dd\rho
 =2\re\int_0^a\overline{u(\rho,\theta)}\,\partial_\rho u(\rho,\theta)\dd\rho.
\]
Using $\re(\overline u\,\partial_\rho u)\le |u|\,|\partial_\rho u|$ and
$2AB\le A^2+B^2$ with $A=|u|/\sqrt{\rho}$ and
$B=\sqrt{\rho}\,|\partial_\rho u|$ gives the first inequality below.
For the second, apply \eqref{eq:angular_mean_van} to
$f_\rho=u(\rho,\cdot)$ at each $\rho$ and integrate against $\dd\rho/\rho$
over $(0,a)$.
\[
 \begin{aligned}
 \int_0^{2\pi}|u(a,\theta)|^2\dd\theta
 &\le\int_0^{2\pi}\int_0^a
       \left(\frac{|u|^2}{\rho^2}+|\partial_\rho u|^2\right)
       \rho\,\dd\rho\dd\theta\\
 &\le\int_0^{2\pi}\int_0^a
       \left(\frac{|\partial_\theta u|^2}{\rho^2}+|\partial_\rho u|^2\right)
       \rho\,\dd\rho\dd\theta
 =\int_{B_a}|\nabla_r u|^2\dd r.
 \end{aligned}
\]
Inserting this into \eqref{eq:nn-outer-hardy} and adding the Hardy estimate \eqref{eq:nn-disk-hardy} on $B_a$ yields
\[
 \begin{aligned}
 \int_\Omega\frac{|u|^2}{|r|^2}\dd r
 &\le \left(1+\tfrac43\log3\right)\int_{B_a}|\nabla_r u|^2\dd r
       +2(\log3)^2\int_{\Omega\setminus B_a}|\partial_\rho u|^2\dd r\\
 &\le C_*\int_\Omega|\nabla_r u|^2\dd r,
 \end{aligned}
\]
where $C_*=\max\{1+\tfrac43\log3,\,2(\log3)^2\}<\tfrac83$. Indeed, Cauchy--Schwarz on $[1,3]$ gives
$(\log3)^2<2\int_1^3 x^{-2}\dd x=\tfrac43$, and hence $\log3<\tfrac54$. For $N=2$ we have  $\Omega=\R^2$. Taking $s\to\infty$ in \eqref{eq:nn-disk-hardy} gives the same estimate with constant $1\le C_*$ by monotone convergence.

Integrating out the remaining positions directly gives
\[
 \int_{D_{ij}}\frac{|\Psi(X)|^2}{|x_i-x_j|^2}\dd X
 \le \frac{C_*}{4}
      \int_{D_{ij}}|\nabla_i\Psi-\nabla_j\Psi|^2\dd X.
\]
By \eqref{eq:nn-decomposition} and
$|\nabla_i\Psi-\nabla_j\Psi|^2\le
2(|\nabla_i\Psi|^2+|\nabla_j\Psi|^2)$, summing over ordered pairs gives
\[
 \begin{aligned}
 \sum_{i=1}^N\int\frac{|\Psi(X)|^2}{d_i(X)^2}\dd X
 &\le\frac{C_*}{2}\sum_{i\ne j}
    \int_{D_{ij}}(|\nabla_i\Psi|^2+|\nabla_j\Psi|^2)\dd X\\
 &\le 3C_*\,T[\Psi]\le8\,T[\Psi].
 \end{aligned}
\]
For the last estimate, the terms with $|\nabla_i\Psi|^2$ are counted
once for almost every $X$ by (i). The terms with $|\nabla_j\Psi|^2$
are counted at most five times by \eqref{eq:nn-multiplicity}.

As in the proof of Theorem~\ref{thm:sharp} we can approximate a $\Psi$ by a general antisymmetric $\Psi\in\dot H^1(\R^{2N})$.
\end{proof}

\subsection{High occupancies}
In this subsection we prove Lemma~\ref{lem:grid-high}, which bounds the contribution to the potential from squares of high occupancy. We first derive a weighted local kinetic-energy estimate from the Neumann eigenvalue sums on a square. We then choose weights in that sum that control the high-occupancy sum.

\begin{proof}[Proof of Lemma~\ref{lem:grid-high}]
Fix $M=16m$ for some $m\in\N$ with $m>32\pi$. Write $n_Q=n_Q(X)$.
For $Q\in\Gamma_k$ with $k\ge1$, let $\widehat Q\in\Gamma_{k-1}$ be the
unique square containing $Q$. If $N\le M$, the high-occupancy sum
vanishes. We assume $N>M$ below.

\paragraph{\textbf{Weighted local Neumann estimate}}
Let $\omega_Q(X)$ be any nonnegative weight depending only on $n_Q$ and
$n_{\widehat Q}$ and vanishing when $n_Q\le m$. With
$B(n)=2\pi(1-\sqrt{8\pi/n})$ for $n\ge1$, we first establish
\begin{equation}\label{eq:weighted-square-neumann}
 \int\omega_Q B(n_Q)\frac{n_Q^2}{|Q|}|\Psi|^2\dd X
 \le\int\omega_Q\sum_{i:x_i\in Q}|\nabla_i\Psi|^2\dd X,
\end{equation}
with the integrand on the left set to zero where $\omega_Q=0$.

Denote by $\lambda_j(Q)$ the Neumann eigenvalues on $Q$, listed in
nondecreasing order and counted with multiplicity. For $n\ge1$, let
$e_n(Q)=\sum_{j=1}^n\lambda_j(Q)$.

Fix the labels of the $n\ge1$ particles in $Q$, and write $X_Q\in Q^n$
for their coordinates and $Y\in(Q^c)^{N-n}$ for the remaining
coordinates, with the original labels understood. For fixed $Y$,
$\Psi(X_Q,Y)$ is antisymmetric in $X_Q$. The Slater determinants of
one-particle Neumann eigenfunctions form an eigenbasis of the
Neumann Laplacian on the antisymmetric subspace of $L^2(Q^n)$, with
lowest eigenvalue $e_n(Q)$. Hence
\begin{equation}\label{eq:neuman_ineq}
 \int_{Q^n}\sum_{i:x_i\in Q}|\nabla_i\Psi(X_Q,Y)|^2\dd X_Q
 \ge e_n(Q)\int_{Q^n}|\Psi(X_Q,Y)|^2\dd X_Q.
\end{equation}

The calculation in Appendix~\ref{app:neumann} gives
\begin{equation}\label{eq:neumann}
 e_n(Q)\ge \frac{2\pi n^2-4\pi\sqrt{2\pi}\,n^{3/2}}{|Q|}
          =B(n)\frac{n^2}{|Q|}.
\end{equation}

For fixed $Y$ and fixed particle labels in $Q$, the occupancies $n_Q$
and $n_{\widehat Q}$ are constant in $X_Q$, and so is $\omega_Q$.
Multiply \eqref{eq:neuman_ineq} by $\omega_Q$ and apply \eqref{eq:neumann}.
Integrating over $Y$ and summing over the choices of particle labels
in $Q$ gives \eqref{eq:weighted-square-neumann}.

\paragraph{\textbf{Choosing the weights}}
For $k\ge1$, let $Q_{i,k}$ be the unique square in $\Gamma_k$ containing
$x_i$. Summing \eqref{eq:weighted-square-neumann} over $Q\in\Gamma_k$,
$k\ge1$, and using Tonelli's theorem gives
\[
 \int |\Psi|^2\sum_{k\ge1}\sum_{Q\in\Gamma_k}
       \omega_Q B(n_Q)\frac{n_Q^2}{|Q|}\dd X
 \le \sum_{i=1}^N\int |\nabla_i\Psi|^2
       \sum_{k\ge1}\omega_{Q_{i,k}}\dd X .
\]

Thus Lemma~\ref{lem:grid-high} follows if we find weights such that, for
configurations $X$ of distinct particles,
\[
 \begin{aligned}
 \text{i)}\quad &
 (\log4)S_{\ell_0,\mathrm{high}}^{(M)}(X)
 \le \sum_{k\ge1}\sum_{Q\in\Gamma_k}
           \omega_Q(X)B(n_Q)\frac{n_Q^2}{|Q|},\\
 \text{ii)}\quad &
 \sum_{k\ge1}\omega_{Q_{i,k}}(X)\le \frac{\log N+11}{2\pi}
 \qquad (1\le i\le N).
 \end{aligned}
\]

We will show that the following choice satisfies both conditions.
For $Q\in\Gamma_k$ with $k\ge1$, set $\alpha_Q=n_Q/n_{\widehat Q}$
when $n_{\widehat Q}>0$ and $\alpha_Q=0$ otherwise. We choose
\begin{equation}\label{eq:kinetic-weight}
 \omega_Q(X)=
 \begin{cases}
 \displaystyle\frac{1}{B(n_Q)}
    \left(\log\frac{n_{\widehat Q}}{n_Q}+\frac32\1_{\{n_Q\le M\}}\right),
     & n_{\widehat Q}>M,\quad \alpha_Q\ge1/16,\\[6pt]
 0,  & \text{otherwise}.
 \end{cases}
\end{equation}
Note that for $nQ\ge n_{\widehat Q}/16>m$ we have $B(n_Q)>\pi>0$.

\paragraph{\textbf{Proof of (i)}}
Using $n(n-1)\le n^2$, $n\in \N$ in  \eqref{eq:high-occupancy-sum} for each such square, it suffices to prove
\begin{equation}\label{eq:high-comparison}
 (\log4)\sum_{k\ge0}\sum_{\substack{P\in\Gamma_k\\n_P>M}}\frac{n_P^2}{|P|}
 \le \sum_{k\ge1}\sum_{Q\in\Gamma_k}
          \omega_Q B(n_Q)\frac{n_Q^2}{|Q|}.
\end{equation}

Fix a configuration $X$ of distinct points and a square $P\in\Gamma_k$
with $n_P>M$. The four squares $Q\in\Gamma_{k+1}$ contained in $P$
satisfy $\widehat Q=P$ and partition $P$. Thus the fractions
$\alpha_Q=n_Q/n_{\widehat Q}$ sum to one and have mean $1/4$.
Since $|Q|=|P|/4$, for $n_Q>0$ we have
\[
 \frac{n_Q^2}{|Q|}\log\frac{n_P}{n_Q}
 =4\frac{n_P^2}{|P|}\,\alpha_Q^2\log\frac1{\alpha_Q}, \qquad \widehat Q = P.
\]
Define
\[
 g(\alpha)=\alpha^2\log\frac1\alpha\quad(0<\alpha\le1),
 \qquad g(0)=0.
\]
For $0<\alpha\le1$, Taylor's formula at $1/4$ and
$g''(\alpha)=2\log(1/\alpha)-3\ge-3$ give
\[
 \begin{aligned}
 g(\alpha)
 &\ge g(1/4)+g'(1/4)(\alpha-1/4)-\frac32(\alpha-1/4)^2\\
 &=L(\alpha)-\frac32\alpha^2,
 \end{aligned}
\]
where
\[
 L(\alpha)=\frac{\log4+3/2}{16}
        +\frac{\log4+1}{2}\left(\alpha-\frac14\right).
\]
The function $L$ is increasing and satisfies
\[
 L(1/16)=\frac{\log4+3/2}{16}
           -\frac{3}{16}\frac{\log4+1}{2}
         =-\frac{\log4}{32}<0.
\]
Consequently we have
\[
 g(\alpha)\1_{\{\alpha\ge1/16\}}
 \ge L(\alpha)-\frac32\alpha^2\1_{\{\alpha\ge1/16\}}
 \qquad(0\le\alpha\le1).
\]
Choose now $\alpha=\alpha_Q$ and sum this inequality over the four squares with $\widehat Q=P$.
Since $\sum_{\widehat Q=P}(\alpha_Q-1/4)=0$, we have
$\sum_{\widehat Q=P}L(\alpha_Q)=4L(1/4)$.
Rearranging and inserting the value of $L(1/4)$ gives
\[
 \sum_{\substack{\widehat Q=P\\\alpha_Q\ge1/16}}
     \left(g(\alpha_Q)+\frac32\alpha_Q^2\right)
 \ge \frac{\log4+3/2}{4}.
\]
Multiply by $4n_P^2/|P|$. For each square in the sum, $|Q|=|P|/4$
and $\alpha_Q=n_Q/n_P$, so
\[
 4\frac{n_P^2}{|P|}\alpha_Q^2=\frac{n_Q^2}{|Q|},
 \qquad \log\frac1{\alpha_Q}=\log\frac{n_{\widehat Q}}{n_Q}.
\]
Using $g(\alpha_Q)=\alpha_Q^2\log(1/\alpha_Q)$, we then find
\[
 \sum_{\substack{\widehat Q=P\\\alpha_Q\ge1/16}}
     \frac{n_Q^2}{|Q|}\left(\log\frac{n_{\widehat Q}}{n_Q}+\frac32\right)
 \ge \left(\log4+\frac32\right)\frac{n_P^2}{|P|}.
\]
Subtracting $(3/2)n_P^2/|P|$ from both sides gives
\begin{equation}\label{eq:four-child}
 (\log4)\frac{n_P^2}{|P|}
 \le \sum_{\substack{\widehat Q=P\\\alpha_Q\ge1/16}}
           \frac{n_Q^2}{|Q|}\log\frac{n_{\widehat Q}}{n_Q}
    +\frac32\left(\sum_{\substack{\widehat Q=P\\\alpha_Q\ge1/16}}
                  \frac{n_Q^2}{|Q|}
                  -\frac{n_P^2}{|P|}\right).
\end{equation}

We now sum over $P\in\Gamma_k$ with $n_P>M$ and over $k\ge0$. Set
\[
 H_k=\sum_{\substack{P\in\Gamma_k\\n_P>M}}\frac{n_P^2}{|P|}.
\]
Summing the parentheses in \eqref{eq:four-child} at level $k$ gives
\[
 \sum_{\substack{P\in\Gamma_k\\n_P>M}}
 \left(\sum_{\substack{\widehat Q=P\\\alpha_Q\ge1/16}}\frac{n_Q^2}{|Q|}
             -\frac{n_P^2}{|P|}\right)
 \le H_{k+1}-H_k
    +\sum_{\substack{Q\in\Gamma_{k+1}\\n_Q\le M<n_{\widehat Q}\\\alpha_Q\ge1/16}}
       \frac{n_Q^2}{|Q|},
\]
For a configuration of distinct particles, $H_k=0$ for all sufficiently
large $k$. The differences therefore telescope to
\[
 \sum_{k\ge0}(H_{k+1}-H_k)=-H_0\le0.
\]
Summing \eqref{eq:four-child} over all $P$ with $n_P>M$ and all levels,
and dropping the nonpositive term $-\frac32 H_0$, gives
\[
 (\log4)\sum_{k\ge0}\sum_{\substack{P\in\Gamma_k\\n_P>M}}\frac{n_P^2}{|P|}
 \le \sum_{k\ge1}\sum_{\substack{Q\in\Gamma_k\\
                  n_{\widehat Q}>M,\ \alpha_Q\ge1/16}}
       \frac{n_Q^2}{|Q|}
       \left(\log\frac{n_{\widehat Q}}{n_Q}+\frac32\1_{\{n_Q\le M\}}\right).
\]
The indicator accounts for the positive quadratic terms with
$n_Q\le M<n_{\widehat Q}$ that remain after telescoping. The cutoff
$\alpha_Q\ge1/16$ still gives $n_Q>m$ for these squares, so the local
Neumann bound applies to them as well. Comparing to \eqref{eq:kinetic-weight}, the
factor in parentheses equals $\omega_Q B(n_Q)$ on the summation set. This gives \eqref{eq:high-comparison}
and proves (i).

\bigskip

\paragraph{\textbf{Proof of (ii)}}
Fix some $i\in \N$, $i\leq N$ and write $n_{i,k}=n_{Q_{i,k}}$. The squares containing $x_i$ fulfill $1\le n_{i,k}\le n_{i,k-1}\le N$ for $k\ge1$.
At every level with $\omega_{Q_{i,k}}(X)>0$, the conditions in the definition of $\omega_Q$ in
\eqref{eq:kinetic-weight} give
\[
 n_{i,k-1}>M,\qquad \frac{n_{i,k-1}}{n_{i,k}}\le16,
 \qquad n_{i,k}>m.
\]
For $n>m$ it is $\sqrt{8\pi/n}<1/2$ and consequently
\[
 \frac{1}{B(n)}
 =\frac{1}{2\pi}\left(1+\frac{\sqrt{8\pi/n}}{1-\sqrt{8\pi/n}}\right)
 \le \frac{1}{2\pi}\left(1+\frac{2\sqrt{8\pi}}{\sqrt n}\right).
\]
To sum the logarithmic terms in \eqref{eq:kinetic-weight}, first use the
term $1/(2\pi)$ in this bound. Each logarithm is nonnegative, so we can
include all levels in the sum. For every $K\ge1$,
\[
 \sum_{k=1}^K\log\frac{n_{i,k-1}}{n_{i,k}}
 =\log\frac{n_{i,0}}{n_{i,K}}\le\log N.
\]
Letting $K\to\infty$ gives a contribution of at most $(\log N)/(2\pi)$. For the second term in the bound on $1/B(n)$, use the ratio bound at every level with $\omega_{Q_{i,k}}(X)>0$ to obtain
\[
 \frac{\log(n_{i,k-1}/n_{i,k})}{\sqrt{n_{i,k}}}
 =\int_{n_{i,k}}^{n_{i,k-1}}\sqrt{\frac{t}{n_{i,k}}}
        \frac{\dd t}{t^{3/2}}
 \le 4\int_{n_{i,k}}^{n_{i,k-1}} t^{-3/2}\dd t.
\]
Indeed, $t\le n_{i,k-1}\le16 n_{i,k}$ on the interval of integration.
Since the occupancies are nonincreasing, these intervals have disjoint interiors, and each lies in $(m,\infty)$. Thus the total contribution of the second term is bounded by
\[
 \frac{2\sqrt{8\pi}}{2\pi}\,4\int_m^\infty t^{-3/2}\dd t
 =\frac{16\sqrt{8\pi/m}}{2\pi}<\frac{8}{2\pi}.
\]

The indicator in $\omega_{Q_{i,k}}$ can contribute only when $n_{i,k}\le M<n_{i,k-1}$. Once $n_{i,k}\le M$, the occupancies at all finer levels are also at most $M$, so this occurs at most once. At such a level with positive weight, $B(n_{i,k})>\pi$, and the indicator term contributes at most $3/(2\pi)$. Combining the three bounds gives
\begin{equation}\label{eq:weight-budget}
 \sum_{k\ge1}\omega_{Q_{i,k}}(X)
 \le\frac{\log N}{2\pi}+\frac{8}{2\pi}+\frac{3}{2\pi}
 =\frac{\log N+11}{2\pi}.
\end{equation}

This proves (ii) and completes the proof of \eqref{eq:grid-high}.
\end{proof}

\appendix
\section{Proof of the square representation}
\label{app:square-representation}

\begin{proof}[Proof of \eqref{eq:square-rep}]
Fix a pair $x_i\ne x_j$, and let $r=|x_i-x_j|$. For fixed $\ell$ and $\theta$, put $v=\ell^{-1}R_{-\theta}(x_i-x_j)$. In coordinate $a\in\{1,2\}$, the separation in the rotated frame is $\ell|v_a|$. If $|v_a|<1$, the points lie in the same grid interval precisely when the smaller coordinate lies in the initial segment of length $\ell(1-|v_a|)$ of its interval. Its position in the interval is uniform under the normalized shift $dz_a/\ell$ on $[0,\ell)$. If $|v_a|\ge1$, no interval contains both coordinates. The shifts in the two coordinates are independent, so the normalized shift measure for which both points lie in the same square is
\[
 A(v)=(1-|v_1|)_+(1-|v_2|)_+.
\]
Averaging over the rotation gives $P(r/\ell)$, where
\[
 P(t)=\frac1{2\pi}\int_0^{2\pi}
                A(t\cos\theta,t\sin\theta)\,d\theta.
\]
With $t=r/\ell$, polar coordinates and $\int_{\R^2}A(v)\,dv=1$ yield
\[
 \int_0^\infty\ell^{-3}P(r/\ell)\,d\ell
 =\frac1{r^2}\int_0^\infty tP(t)\,dt
 =\frac1{2\pi r^2}.
\]
The sum $\sum_Q n_Q(X)(n_Q(X)-1)$ counts each unordered pair in the same square twice. By Tonelli's theorem, we may integrate the nonnegative pair contributions separately. The prefactor $\pi$ in \eqref{eq:square-rep} then gives $2\pi/(2\pi r^2)=r^{-2}$ for each unordered pair, as required.
\end{proof}

\section{Dyadic representation of the truncated potential}
\label{app:truncated-representation}

\begin{proof}[Proof of Lemma \ref{lem:truncated-representation}]
Fix $X$, $\ell_0>0$, $\theta\in[0,2\pi)$, and $k\ge0$. Use the grid
formula at side length $\ell_k=2^{-k}\ell_0$ for all $z\in\R^2$. For every
$t\in\mathbb Z^2$,
\[
 \sum_{Q\in\Gamma(\ell_k,\theta,z)}n_Q(X)(n_Q(X)-1)
 =\sum_{Q\in\Gamma(\ell_k,\theta,z+\ell_k t)}n_Q(X)(n_Q(X)-1).
\]
The square indexed by $j$ at shift $z+\ell_k t$ is the square indexed
by $j+t$ at shift $z$.

For each fixed $t$, integrate this equality over $z\in[0,\ell_k)^2$.
In the shifted integral, set $u=z+\ell_k t$, so that $du=dz$ and
$u\in\ell_k t+[0,\ell_k)^2$. This gives
 \begin{equation}\label{app:eqB1}
 \begin{split}
    &\int_{[0,\ell_k)^2}\sum_{Q\in\Gamma(\ell_k,\theta,z)}
                 n_Q(X)(n_Q(X)-1)\dd z\\
 &\qquad=\int_{[0,\ell_k)^2}\sum_{Q\in\Gamma(\ell_k,\theta,z+\ell_k t)}
                 n_Q(X)(n_Q(X)-1)\dd z\\
 &\qquad=\int_{\ell_k t+[0,\ell_k)^2}\sum_{Q\in\Gamma(\ell_k,\theta,u)}
                 n_Q(X)(n_Q(X)-1)\dd u.
 \end{split}
 \end{equation}
Now sum over $t\in\{0,\ldots,2^k-1\}^2$. On the left we obtain $4^k$
copies of the same integral. The domains on the right satisfy
\[
 \bigsqcup_{t\in\{0,\ldots,2^k-1\}^2}
   \bigl(\ell_k t+[0,\ell_k)^2\bigr)
 = [0,2^k\ell_k)^2=[0,\ell_0)^2.
\]
Dividing \eqref{app:eqB1} by $\ell_0^2=4^k\ell_k^2$ we arrive at
\[
 \begin{aligned}
&\int_{[0,\ell_k)^2}\sum_{Q\in\Gamma(\ell_k,\theta,z)}
       n_Q(X)(n_Q(X)-1)\,\frac{dz}{\ell_k^2}\\
 &\qquad=\int_{[0,\ell_0)^2}\sum_{Q\in\Gamma(2^{-k}\ell_0,\theta,z)}
       n_Q(X)(n_Q(X)-1)\,\frac{dz}{\ell_0^2}.
 \end{aligned}
\]

Split $(0,L]$ into the intervals $(2^{-k-1}L,2^{-k}L]$, $k\ge0$.
On the $k$th interval in \eqref{eq:truncated-potential}, set
$\ell=\ell_k=2^{-k}\ell_0$. Then $L/2<\ell_0\le L$ and
\[
 \frac{d\ell}{\ell^3}
 =\frac{2^{-k}\,d\ell_0}{(2^{-k}\ell_0)^3}
 =\frac1{(2^{-k}\ell_0)^2}\frac{d\ell_0}{\ell_0}.
\]

At fixed $k$, the grid in the sum varies with
$\ell_0$ through its side length $2^{-k}\ell_0$. Integrating over
$L/2<\ell_0\le L$ and $\theta\in[0,2\pi)$ with the factor
$(2^{-k}\ell_0)^{-2}\,\frac{d\ell_0}{\ell_0}\frac{d\theta}{2\pi}$
from the change of scale gives
\[
 \begin{aligned}
 &\int_{2^{-k-1}L<\ell\le 2^{-k}L}
    \sum_{Q\in\Gamma(\ell,\theta,z)}n_Q(X)(n_Q(X)-1)\,
       d\mu(\ell,\theta,z)\\
 &\qquad=\int_{L/2<\ell_0\le L}
    \sum_{Q\in\Gamma(2^{-k}\ell_0,\theta,z)}
       \frac{n_Q(X)(n_Q(X)-1)}{(2^{-k}\ell_0)^2}\,
       \ell_0^2\dd\mu(\ell_0,\theta,z).
 \end{aligned}
\]
In the sum on the right, $|Q|=(2^{-k}\ell_0)^2$. Summing over $k$
and applying Tonelli's theorem and \eqref{eq:dyadic-sum} proves
\eqref{eq:grid-average}.
\end{proof}

\section{Neumann eigenvalue sums on a square}
\label{app:neumann}

\begin{proof}[Proof of \eqref{eq:neumann}]
On the unit square the Neumann eigenvalues are
$\pi^2(k_1^2+k_2^2)$ for $k_1,k_2\in\mathbb N_0$. The lattice points corresponding to eigenvalues at most $\Lambda$ have distance at most $\sqrt{\Lambda}/\pi$ from the origin. The unit cells extending in the positive coordinate directions from these points have disjoint interiors and lie in the quarter-disc of radius
$\sqrt{\Lambda}/\pi+\sqrt2$. Thus for the $j$th eigenvalue on the unit square,
\[
 j\le\frac\pi4\left(\frac{\sqrt{\lambda_j}}{\pi}+\sqrt2\right)^2,
 \qquad
 \lambda_j\ge(2\sqrt{\pi j}-\pi\sqrt2)_+^2
             \ge4\pi j-4\pi\sqrt{2\pi j}.
\]
Summing this inequality, we use $\sum_{j=1}^n j=n(n+1)/2$ and
$\sum_{j=1}^n\sqrt j\le n\sqrt n$. The eigenvalues on $Q$ are those on
the unit square multiplied by $1/|Q|$. This gives \eqref{eq:neumann}.
\end{proof} 
\subsection*{Use of AI}
The authors used AI to explore conjectures and possible proof strategies concerning the present topic. An initial argument for the sharp constant was found with assistance of AI. The resulting arguments were developed and verified by the author.
This version of the draft was checked and revised with AI tools to the extend of language checks, structure of arguments and presentation. The author takes responsibility for the proof and the content of this paper.
\subsection*{Funding}
Support was provided by the European Research Council (ERC) under the ERC Advanced Grant MathBEC (grant agreement No. 101095820), and from the VILLUM Foundation (grant No. VIL73411).


\begin{thebibliography}{9}

\bibitem{HOHLT}
M.~Hoffmann-Ostenhof, T.~Hoffmann-Ostenhof, A.~Laptev, and J.~Tidblom,
\emph{Many-particle Hardy inequalities},
J. Lond. Math. Soc. (2) \textbf{77} (2008), no.~1, 99--115.
\href{https://doi.org/10.1112/jlms/jdm091}{doi:10.1112/jlms/jdm091}.

\bibitem{FHOLS}
R.~L.~Frank, T.~Hoffmann-Ostenhof, A.~Laptev, and J.~P.~Solovej,
\emph{Hardy inequalities for large fermionic systems},
J. Spectr. Theory \textbf{14} (2024), 805--835.
\href{https://arxiv.org/abs/2403.12640}{arXiv:2403.12640}.

\bibitem{FdL}
C.~Fefferman and R.~de la Llave,
\emph{Relativistic Stability of Matter - I},
Rev. Mat. Iberoam. \textbf{2} (1986), no.~2, 119--213.
\href{https://doi.org/10.4171/RMI/30}{doi:10.4171/RMI/30}.

\bibitem{HS}
C.~Hainzl and R.~Seiringer,
\emph{General decomposition of radial functions on $\R^n$ and applications
to $N$-body quantum systems},
Lett. Math. Phys. \textbf{61} (2002), 75--84.
\href{https://arxiv.org/abs/math-ph/0107011}{arXiv:math-ph/0107011}.

\bibitem{LiebYau}
E.~H.~Lieb and H.-T.~Yau,
\emph{The Chandrasekhar theory of stellar collapse as the limit of quantum mechanics},
Comm. Math. Phys. \textbf{112} (1987), no.~1, 147--174.
\href{https://doi.org/10.1007/BF01217684}{doi:10.1007/BF01217684}.

\bibitem{HSVlines}
D.~Hundertmark, M.~R.~Schulz, and S.~Vugalter,
\emph{Why a system of three bosons on separate lines can not exhibit the confinement induced Efimov effect},
arxiv Preprint.
\href{https://arxiv.org/abs/2411.19263}{arXiv:2411.19263}.

\end{thebibliography}
\end{document}